\documentclass[aps,prb,twocolumn,showpacs,superscriptaddress]{revtex4}

\usepackage{graphicx}
\usepackage{amsmath}

\begin{document}

\title{Observation of a $C$-type short range antiferromagnetic order in expanded FeS}
\author{Meng Wang}
\email{wangmeng5@mail.sysu.edu.cn}
\affiliation{School of Physics, Sun Yat-Sen University, Guangzhou, Guangdong 510275, China }
\affiliation{Department of Physics, University of California, Berkeley, California 94720, USA }
\author{Ming Yi}
\email{mingyi@rice.edu}
\affiliation{Department of Physics and Astronomy, Rice University, Houston, TX 77005, USA}
\affiliation{Department of Physics, University of California, Berkeley, California 94720, USA }
\author{Benjamin A. Frandsen}
\affiliation{Department of Physics and Astronomy, Brigham Young University, Provo, Utah 84602, USA}
\affiliation{Materials Science Division, Lawrence Berkeley National Laboratory, Berkeley, California 94720, USA }
\author{Junjie Yin}
\affiliation{School of Physics, Sun Yat-Sen University, Guangzhou, Guangdong 510275, China }
\author{Hualei Sun}
\affiliation{School of Physics, Sun Yat-Sen University, Guangzhou, Guangdong 510275, China }
\author{Zhijun Xu}
\affiliation{NIST Center for Neutron Research, National Institute of Standards and Technology, Gaithersburg, Maryland 20899, USA }
\affiliation{Department of Materials Science and Engineering, University of Maryland, College Park, Maryland 20742, USA }
\author{Huibo Cao}
\affiliation{Neutron Scattering Division, Oak Ridge National Laboratory, Oak Ridge, Tennessee 37831, USA}
\author{Edith Bourret-Courchesne}
\affiliation{Materials Science Division, Lawrence Berkeley National Laboratory, Berkeley, California 94720, USA }
\author{Jeffrey W. Lynn}
\affiliation{NIST Center for Neutron Research, National Institute of Standards and Technology, Gaithersburg, Maryland 20899, USA }
\author{Robert J. Birgeneau}
\affiliation{Department of Physics, University of California, Berkeley, California 94720, USA }
\affiliation{Materials Science Division, Lawrence Berkeley National Laboratory, Berkeley, California 94720, USA }

\begin{abstract}

We report neutron diffraction studies of FeS single crystals obtained from Rb$_x$Fe$_{2-y}$S$_2$ single crystals via a hydrothermal method. While no $\sqrt {5}\times \sqrt {5}$ iron vacancy order or block antiferromagnetic order typical of Rb$_x$Fe$_{2-y}$S$_2$ is found in our samples, we observe $C$-type short range antiferromagnetic order with moments pointed along the $c$-axis hosted by a new phase of FeS with an expanded inter-layer spacing. The N\'{e}el temperature for this magnetic order is determined to be 165 K. Our finding of a variant FeS structure hosting this $C$-type antiferromagnetic order demonstrates that the known FeS phase synthesized in this method is in the vicinity of a magnetically ordered ground state, providing insights into understanding a variety of phenomena observed in FeS and the related FeSe$_{1-x}$S$_x$ iron chalcogenide system.

\end{abstract}

\maketitle
\section{Introduction}

High temperature superconductivity (HTSC) has always been found close to magnetism. Specifically, most of the parent compounds of the copper- and iron-based superconductors exhibit antiferromagnetic (AF) orders\cite{Lynn2009}. In the iron-based superconductors, the AF order is typically accompanied by a tetragonal to orthorhombic ($T-O$) structural transition. Upon carrier doping, isovalent substitution, or pressure, these competing phases are suppressed and superconductivity emerges\cite{Johnston2010,Si2016}.

FeSe, the iron-based superconductor with arguably the simplest structure, appears to differ in that it does not show long-range magnetic order at ambient pressure, while still hosting a robust electronic nematic phase marked by a relatively high $T-O$ structural transition \cite{Hsu2008,McQueen2009,Fernandes2014,Shimojima2014,Nakayama2014,Baek2014,Watson2015a,Wen2016,Wang2016a,Kothapalli2016,Yi2019}.
Interestingly, magnetic order is shown to appear under hydrostatic pressure by muon spin rotation ($\mu$SR) and transport measurements between $2\sim6$ GPa\cite{Bendele2012,Sun2015,Khasanov2017}. Meanwhile, superconductivity is also enhanced under pressure, reaching a maximum T$_c\approx38$ K at a pressure of 6 GPa, where both the nematic and magnetic orders are absent\cite{Sun2015}. Motivated by this pressure-induced change of the magnetic and superconducting properties, the chemical pressure phase diagram by way of substitution of Se by S has also been explored. While it is shown that substitution of Se by S suppresses the electronic nematicity, no magnetic order is found nor is T$_c$ enhanced in FeSe$_{1-x}$S$_x$\cite{Watson2015,Hosoi2016,Xiang2017,Matsuura2017}, with T$_c$ limited to $\sim4$ K at the FeS end of the phase diagram\cite{Lai2015}. One possible explanation for the lack of T$_c$ enhancement is the decrease of the electronic correlations with the increase of S substitution\cite{Reiss2017,Yi2015}. Although FeS appears far away from the $T-O$ structural transition and electronic nematic order as demonstrated by extensive experimental results, the possibility of magnetic instabilities in the vicinity of FeS has not been fully explored. Neutron scattering experiments on FeS have revealed magnetic excitations similar to those in other iron-based materials, as well as features reminiscent of magnetic Bragg peaks\cite{Kuhn2017,Man2017}, while $\mu$SR measurements suggest the existence of weak disordered magnetism\cite{Holenstein2016}.

\begin{figure}[t]
\includegraphics[scale=0.5]{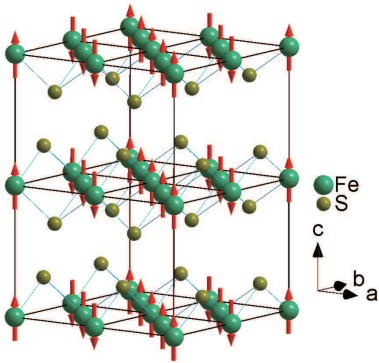}
\caption{ Proposed $C$-type AF order with moments aligned along the $c$ axis for expanded FeS. The distance between two iron layers of the expanded FeS is 5.92 \AA\ in-between 5.09 \AA\ for that of reported FeS  and 6.95 \AA\ for that of Rb$_x$Fe$_{2-y}$S$_2$.  }
\label{structure}
\end{figure}

In this paper, we use neutron diffraction to explore the magnetic phase space of single crystals of FeS synthesized from Rb$_x$Fe$_{2-y}$S$_2$ using a hydrothermal method\cite{Lin2016}. While no magnetic peaks corresponding to the commonly studied FeS structure are observed at the probed wave vectors, a new FeS structure with an expanded inter-layer spacing is identified and found to host short-range $C$-type AF order, as shown in Fig. \ref{structure}. The moments point along the $c$-axis, similar to that of the known magnetic order in Rb$_x$Fe$_{2-y}$S$_2$\cite{Wangm2014,Wang2016b}. 

\section{Experimental Details}

FeS single crystals were obtained from Rb$_x$Fe$_{2-y}$S$_2$ using a hydrothermal method that was developed previously\cite{Lin2016}. Energy-dispersive x-ray spectroscopy (EDX) was employed to determine the composition of the samples. The energy of the electron beam was fixed at 20.0 KeV. Single crystals and powders grounded from the single crystal samples were characterized by a DX-27 mini x-ray diffractometer with Cu $K\alpha$ radiation. Our neutron diffraction experiments were carried out on both the HB3A four-circle spectrometer at the High Flux Isotope Reactor, Oak Ridge National laboratory (HFIR, ORNL) and the BT-7 thermal triple-axis spectrometer at the NIST Center for Neutron Research (NCNR, NIST)\cite{Lynn2012}. Two single crystals weighing 10 and 40 mg were measured on HB3A and BT-7, respectively. Throughout this work, we define the magnitude of the momentum transfer as $|Q|=2\pi\sqrt{(H/a)^2+(K/b)^2+(L/c)^2}$, where $(H, K, L)$ are the Miller indices in reciprocal lattice units (r.l.u), and the lattice constants are $a=b=3.68$ \AA\, and $c=5.09$ \AA\ as determined from the diffraction experiments.

\section{Results}

\begin{figure}[t]
\includegraphics[scale=0.8]{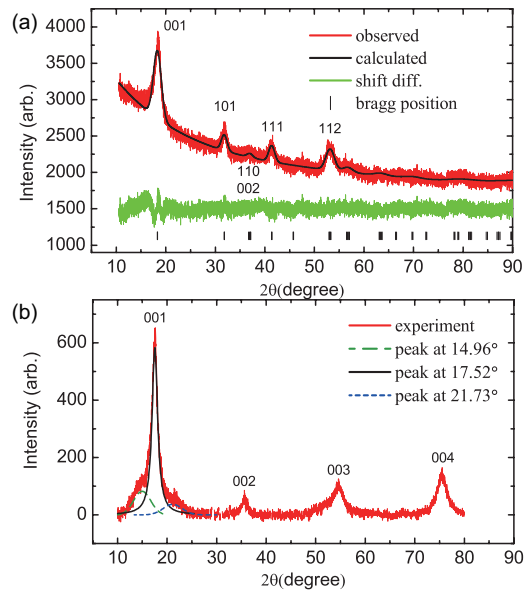}
\caption{ (a) XRD  measurement on powder samples of FeS at room temperature. The solid line is a Ritveld refinement profile with the commonly observed lattice constants $a=b=3.68$ \AA\, and $c=5.09$ \AA. (b) X-ray diffraction pattern of a FeS single crystal showing nuclear reflection peaks of $Q=(0, 0, L), L=1, 2, 3$, and 4. In addition to the main peak of  $Q=(0, 0, 1)$ at 17.52$^\circ$, a broad peak at 14.96$^\circ$ and a weak peak at 21.73$^\circ$ exist. The lines are fits to a Gaussian function throughout this paper.}
\label{fig0}
\end{figure}

The composition of our sample was determined to be FeS$_x$ ($x=0.975\pm0.015$) from EDX. The content of the alkali metal Rb is not detectable. X-ray diffraction (XRD) measurements were conducted on both a powder sample and a single crystal sample to investigate the structure of FeS. In addition to the expected peaks of FeS in Fig. \ref{fig0}, there are two weak shoulders on two sides of the $(0, 0, 1)$ nuclear peak, indicating the existence of the other phases.

\begin{figure}[t]
\includegraphics[scale=0.4]{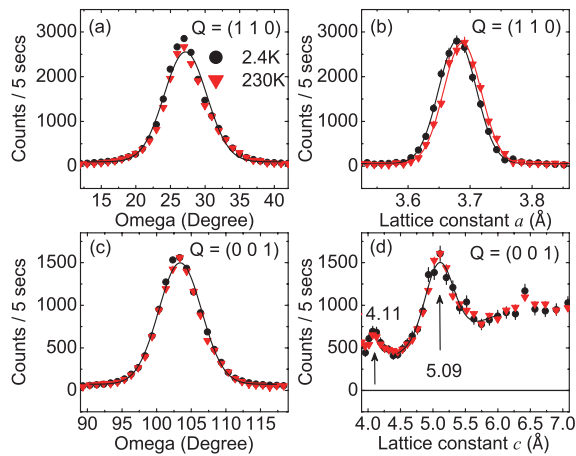}
\caption{ (a) Rocking curve scans of a nuclear peak at $Q=(1, 1, 0)$ , $T=2.4$ and 230 K, respectively.  (b) The lattice constant $a$ in units of \AA\ translated from $\theta-2\theta$ scans at $Q=(1, 1, 0)$, $T=2.4$ and 230 K.  (c) Similar rocking curve scans at $Q=(0, 0, 1)$ and (d) the lattice constant $c$ translated from $\theta-2\theta$ scans at  $Q=(0, 0, 1)$. Identical scans at 2.4 and 230 K reveal that no significant structural deformation occurs at 230 K. The error bars stand for one standard deviation of the measured counts throughout this paper. }
\label{fig1}
\end{figure}

 \begin{figure}[t]
\includegraphics[scale=0.4]{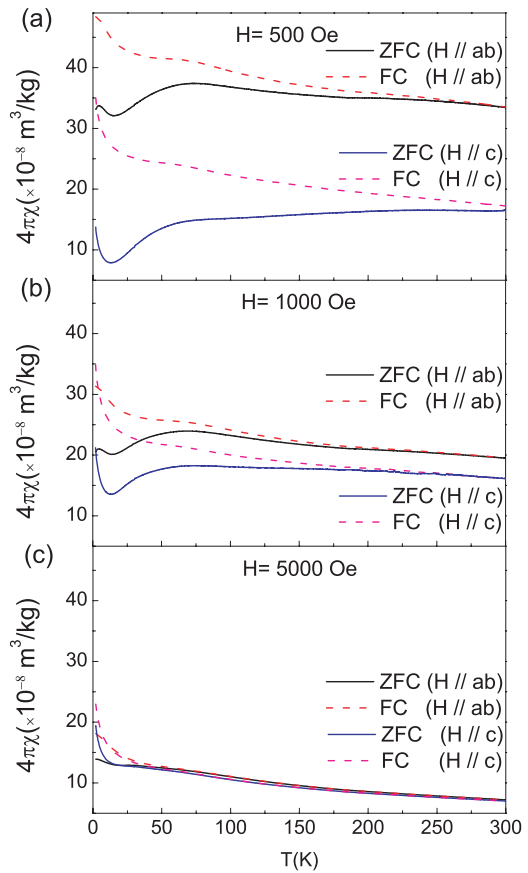}
\caption{ Magnetic susceptibility measurements for a sample weighing 5.6 mg with fields of (a) $H=$500, (b) 1000, and (c) 5000 Oe parallel to the $ab$-plane and the $c$-axis of FeS. ZFC is zero field cooled. FC is field cooled. }
\label{sus}
\end{figure}

\begin{figure*}[t]
\includegraphics[scale=0.6]{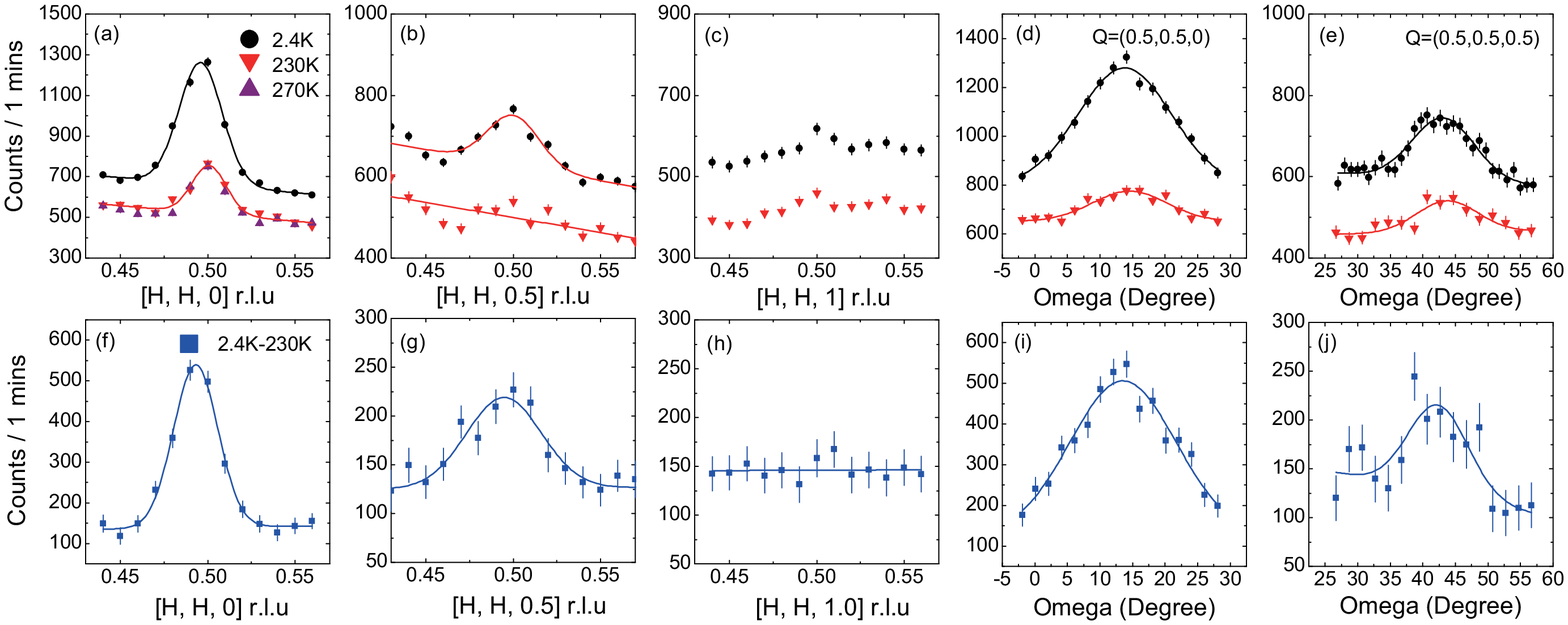}
\caption{ (a) Magnetic Bragg peak scans along the $[H, H]$ direction at $Q=(0.5, 0.5, 0)$, (b) $Q=(0.5, 0.5, 0.5)$, and (c) $Q=(0.5, 0.5, 1)$ at 2.4, 230, and 270 K. (d) Rocking curve scans at $Q=(0.5, 0.5, 0)$ and (e) $Q=(0.5, 0.5, 0.5)$. (f-j) Intensity differences between 2.4 K and 230 K corresponding to the scans in (a-e). The counts are normalized from data collected for 2 or 3 minutes per point for statistics.}
\label{fig2}
\end{figure*}

We then investigate the structure of the FeS crystals produced in this hydrothermal reaction using neutron diffraction. 
During the hydrothermal reaction which removes the intercalated Rb atoms from Rb$_x$Fe$_{2-y}$S$_2$ single crystals, the crystallographic $c$-axis is collapsed from 13.9\AA\ in Rb$_x$Fe$_{2-y}$S$_2$~\cite{Wang2015b} to 5.03\AA\ in FeS~\cite{Lai2015,Lin2016}. This large change is likely to result in significant mosaicity of the sample along the $c$-axis direction. From a comparison of the nuclear Bragg peaks at $Q=(1, 1, 0)$ and $Q=(0, 0, 1)$ measured at 2.4 K and 230 K (Fig.~\ref{fig1}), we see no structural distortion of our single crystal at low temperature. Fits to the rocking curves in Figs.~\ref{fig1}(a, c) yield an identical mosaicity of $7^\circ$ within errors.  The lattice constants are extracted from $\theta-2\theta$ scans, which have been converted into $d$-spacings in \AA\ according to Bragg's law (Figs. \ref{fig1}(b, d)). The $\theta-2\theta$ scan at $Q=(1, 1, 0))$ reveals an in-plane lattice constant of $a=b=3.68$ \AA\ at 2.4 K (3.69 \AA\ at 230 K).  However, the scan at $Q=(0, 0, 1))$ shows a prominent peak and a smaller secondary peak, corresponding to $c=5.09\pm0.02$ and $4.11\pm0.03$  \AA, respectively. In addition, there is a very broad feature that corresponds to a continuous spread of $c$-axis spacings as shown in Fig. \ref{fig1}(d). The prominent peak corresponds to a portion of the sample with $c=5.09\pm0.02$ \AA, consistent with previous reports of FeS\cite{Lai2015,Borg2015,Lin2016,Kuhn2017}. 
We notice that the smaller peak at $c=4.11\pm0.03$ \AA\ corresponds to $Q=1.53\pm0.01$ \AA$^{-1}$, which is the same as that of the right shoulder at $21.73^\circ$ observed in the single crystal XRD spectrum in Fig. \ref{fig0}(b). 
This peak at $Q=1.53\pm0.01$ \AA$^{-1}$ is consistent with the prominent (0, 2, 0) peak of thiourea (CH$_4$N$_2$S)\cite{Begum2009}, which is used in the hydrothermal synthesis. We therefore attribute the peak at $Q = 1.53$ \AA$^{-1}$ to thiourea, rather than to a phase of FeS with a compressed layer spacing of 4.11 \AA.
The broad shoulder on left at $14.96^\circ$ in Fig. \ref{fig0}(b) corresponds to a layer spacing of approximately 5.92 \AA\, which falls within the continuous spread of $c$-axis spacings observed in neutron diffraction measurements in Fig. \ref{fig1}(d). 
We note that the distance between FeS layers in Rb$_x$Fe$_{2-y}$S$_2$ is 6.95 \AA \cite{Wangm2014,Wang2016b}. Therefore, the observation of a peak at 5.92 \AA\ suggests the existence of an FeS phase with an interlayer distance that is intermediate between standard FeS (5.09 \AA) and Rb$_x$Fe$_{2-y}$S$_2$ (6.95 \AA), which can be accessed via hydrothermal synthesis. We will refer to this phase of FeS with an interlayer spacing of 5.92 \AA\ as "expanded FeS". The expanded interlayer distance of 5.92 \AA\ may be caused by residual thiourea during the hydrothermal process, as indicated by the Bragg peak at $Q=1.53$ \AA$^{-1}$ possibly from thiourea.

In Fig. \ref{sus}, we show field dependent splittings of the magnetic susceptibilities between the field cooled (FC) and zero field cooled (ZFC) measurements\cite{Kuhn2017}. A weak diamagnetic response presents below 4 K for the ZFC measurements with H$\parallel$$ab$-plane, $H=500$ and 1000 Oe. The diamagnetic response is absent for H$\perp$$ab$-plane in Figs. \ref{sus}(a, b) and Fig. \ref{sus}(c) with $H=5000$ Oe. The varied responses are consistent with the appearance of an $in$-plane superconductivity at 4 K in layered FeS\cite{Lai2015}. The enhanced susceptibilities for H$\parallel$$ab$-plane compared to that of H$\perp$$ab$-plane indicate the existence of $c$ axis polarized moments. To check whether the block antiferromagnetic (AF) order and $\sqrt{5}\times\sqrt{5}$ iron vacancy order in Rb$_x$Fe$_{2-y}$S$_2$ exist in the resultant FeS single crystals, we conducted neutron diffraction measurements on the HB3A four-circle diffractometer\cite{Wangm2014,Wang2016b,Chak2014}. No peaks are found at the expected wave vectors of either the block AF order or the iron vacancy order (data not shown). 
In addition, the magnetic order, if present, is likely to be weak. To investigate this potentially weak magnetic order, we carried out diffraction measurements at the BT-7 thermal triple-axis spectrometer at NCNR-NIST. Having known the absence of strong magnetic order and iron vacancy order from preliminary measurements, we aligned a larger single crystal with 40 mg in the $[H, H, L]$ plane to search for evidence of possible weak magnetic order. 

In Fig. \ref{fig2}, we present key results from our investigation of the magnetic order. A clear peak centered on $Q=(0.5, 0.5, 0)$ is present at 2.4~K, shown in Fig. \ref{fig2}(a). Another peak is present at $Q=(0.5, 0.5, 0.5)$ but with a reduced intensity, and no observable peak is found at $Q=(0.5, 0.5, 1)$ as shown in Figs. \ref{fig2}(b, c). We performed rocking curve scans at $Q=(0.5, 0.5, 0)$ and $Q=(0.5, 0.5, 0.5)$, revealing an average full width at the half maximum (FWHM) of 14$^\circ$ seen in Fig. \ref{fig2}(d, e). At a higher temperature of 230~K, a small peak still remains at $Q=(0.5, 0.5, 0)$. However, no change in intensity is observed with a further increase of the temperature up to 270~K, suggesting that the small peak that remains is from nuclear reflections. Curiously, the background is lower at 230 and 270~K than 2.4~K. This possibly results from the large incoherent neutron scattering of H in the thiourea inherited from the hydrothermal synthesis process or related to the weak disordered magnetism suggested by $\mu SR$ measurements\cite{Holenstein2016}. The differences in intensity between 2.4~K and 230~K are plotted in Figs. \ref{fig2}(f-j). Gaussian fits to the peaks at $L=0$ and 0.5 along the $[H, H]$ direction result in FWHMs of 0.03 and 0.05 r.l.u., yielding short-range correlation lengths of 77 and 46 \AA, respectively. The way we estimate the magnetic correlation length has been described elsewhere\cite{Wang2016b}.

\begin{figure}[t]
\includegraphics[scale=0.6]{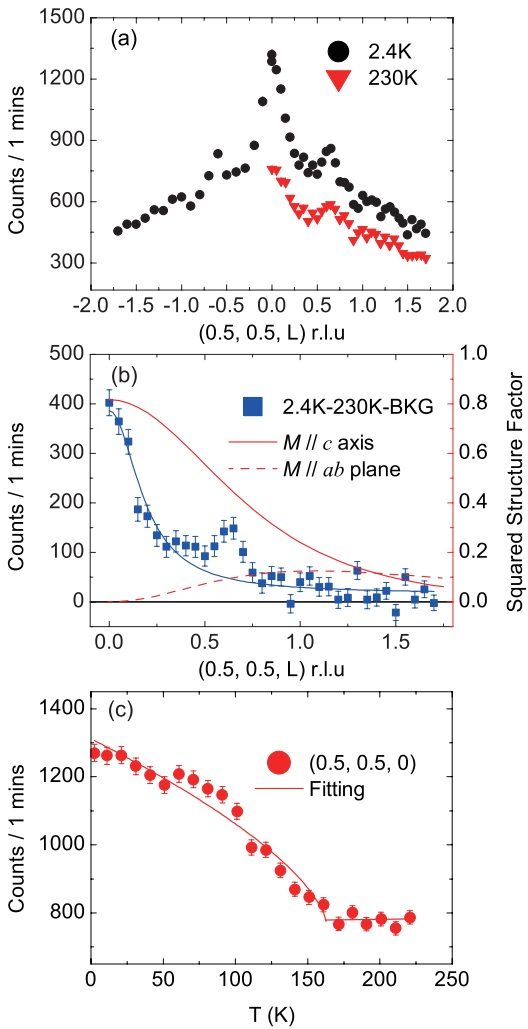}
\caption{ (a) A scan along the $(0.5, 0.5, L)$ direction from $L=-1.75$ to 1.75 at 2.4 K and  a similar scan from $L=0$ to 1.75 at 230 K. (b) Intensity differences between 2.4 K and 230 K of the scans along the $(0.5, 0.5, L)$ direction. The blue curve is a guide simulated by a Lorentz function. The solid and dashed red curves are the squared structure factors corresponding to moments aligned along the $c$ axis or in the $ab$ plane, respectively. The structure factors have included the effects of the magnetic form factor, moment direction, and twinning.  (c) Magnetic order parameters measured at $Q=(0.5, 0.5, 0)$, yielding a N\'{e}el temperature at $T_N=170$ K. The solid curve is a fit to a simple power-law dependence $\phi(T)^2\propto(1-T/T_N)^{2\beta}$.  }
\label{fig3}
\end{figure}

To investigate the arrangement of the moments along the $c$-axis, we present elastic scans along the $(0.5, 0.5, L)$ direction at 2.4 and 230 K in Fig. \ref{fig3}(a). The intensities are symmetric in the range $-1.75\leq L\leq1.75$. We have collected more points in the positive regime to improve statistics. The intensity difference between the two temperatures is plotted in Fig. \ref{fig3}(b). 

The magnetic peak intensity decreases dramatically with increasing $L$, which is due in part to the magnetic form factor, but could also indicate a magnetic order with moments along the $c$-axis. To investigate this further, we compare in Fig.~\ref{fig3}(b) the calculated squared magnetic structure factor $\arrowvert F(Q)\arrowvert^2$ for magnetic moments aligned along the $c$ axis (solid red curve) with that resulting from moments in the $ab$ plane (dashed red curve). The structure factor $\arrowvert F(Q)\arrowvert^2$ is proportional to $\arrowvert f(Q) \cdot sin(\phi)\arrowvert^2$, where $f(Q)$ is the magnetic form factor of Fe$^{2+}$ and $\phi$ is the angle between the moments and the scattering vector. As has been well-studied, the stripe AF order found in LaOFeAs, BaFe$_2$As$_2$, and NaFeAs exhibits magnetic peaks at $Q=(n/2, m/2, p/2)$, where $n, m, p$ are odd (using a magnetic unit cell with a single FeAs layer to be equivalent to FeS studied here)\cite{Lynn2009}. The moments in the stripe AF configuration are in-plane, thus the magnetic peak intensities follow the dashed curve in Fig. \ref{fig3}(b). Clearly, the measured intensities of the peaks as a function of $L$ match better with the squared magnetic structure factor with the moments $M$ parallel to the $c$ axis than parallel in the $ab$ plane. Taking the most prominent peak centered at $L=0$ along the $L$ direction\cite{Bao1997}, the data reveal the existence of a $C$-type antiferromagnetic order as illustrated in Fig. \ref{structure}.  A fit to the widths of the peaks along the $L$ direction reveals a correlation length of $\sim$18 \AA, which is comparable to the length of three unit cells along the $c$ axis. There is a weak peak located at $L=0.63$ in Fig. \ref{fig3} (b), which persists up to 230K, as shown in Fig. \ref{fig3} (a). 
The momentum transfer corresponding to this peak is $Q=1.43$ \AA$^{-1}$, the same as that of the $(1, 0, 1)$ peak of thiourea\cite{Begum2009}. Therefore, this peak is highly likely to be associated with the effect of temperature on the structure of the residual thiourea in the sample.

We determined the magnetic order parameter using the peak intensities at $Q=(0.5, 0.5, 0)$.  A simple power-law dependence $\phi(T)^2\propto(1-T/T_N)^{2\beta}$ was explored to fit the order parameter measured at $L=0$, yielding values of $T_N=170\pm4$ K and $\beta=0.38\pm0.04$, nicely consistent with the $\beta\approx0.36$ expected for a three-dimensional Heisenberg magnetic model\cite{Guillou1980,Campostrini2002,Ko2007}.  

\section{Discussion and Conclusion}

The wave vectors, momentum dependence, and temperature dependence of the observed peaks at $Q=(0.5, 0.5, L)$ demonstrate that they originate from $C$-type short-range AF magnetic order within the FeS phase with the expanded inter-layer spacing of $c=5.92$ \AA, as illustrated in Fig. \ref{structure}. Although AF magnetic orders with moments aligned along the $c$ axis have been reported, for example, Sr$_{0.63}$Na$_{0.37}$Fe$_2$As$_2$\cite{Allred2016}, BaMn$_2$As$_2$\cite{Singh2009} and CaCo$_2$As$_2$\cite{Quirinale2013}, they are not the commonly observed parent compounds of layered iron-based superconductors. Some iron-based materials, for example, K$_{0.8}$Fe$_{1.6}$Se$_2$\cite{Bao2011}, Rb$_{0.8}$Fe$_{1.6}$S$_2$\cite{Wangm2014}, and BaFe$_2$Se$_3$\cite{Caron2011} exhibit out-of-plane moments. However, their structures deviate strongly from the layered iron based superconducting materials. A fit of the same power-law dependence $\phi(T)^2\propto(1-T/T_N)^{2\beta}$ to the magnetic order parameter for BaFe$_2$As$_2$ yields $\beta\approx0.1$, which is near the $\beta=0.125$ expected for a two-dimensional Ising system\cite{Wilson2009,Wilson2010}. The different critical exponents suggest a distinct nature of the magnetism in the expanded FeS compared to the other two-dimensional iron-based AF orders.

The FeS samples used in our experiment are obtained from Rb$_x$Fe$_{2-y}$S$_2$, which has a block AF phase with out-of-plane moments and a $\sqrt{5}\times\sqrt{5}$ iron vacancy order. The hydrothermal synthesis process removes the alkali metal atoms in a solution with NaOH and thiourea, and the iron vacancies are filled with excess iron powder added to the solution. The single crystals of Rb$_x$Fe$_{2-y}$S$_2$ then collapse along the $c$ axis, forming FeS. During this process, regions of the sample may form an intermediate and metastable structure with an expanded layer spacing $c=5.92$ \AA\ and out-of-plane moments inherited from Rb$_x$Fe$_{2-y}$S$_2$. This magnetically ordered metastable structure  with remnant thiourea may be unavoidable for large single crystals. Thus, careful attention must be given to the interpretation of experiments based on a large amount of single crystal samples. We call for more research including complete characterization of the chemical make up of this new phase of FeS. Magnetic peaks associated with the dominant phase of FeS ($c=5.09$ \AA) have not been observed in the $[H, H, L]$ plane.

In summary, we have measured the structure of FeS prepared through hydrothermal reactions with Rb$_x$Fe$_{2-y}$S$_2$ single crystals. Neutron diffraction measurements at the thermal triple-axis spectrometer BT-7 have revealed $C$-type short range AF order hosted by a newly discovered FeS structure with an expanded inter-layer spacing of $c=5.92$ \AA. The distance between FeS layers in the present work is intermediate between these commonly reported FeS and Rb$_x$Fe$_{2-y}$S$_2$, suggesting that this $C$-type AF ordered Fe represents an intermediate and metastable state produced in the hydrothermal reaction. The dominant FeS phase with the lattice constant $c=5.09$ \AA\, on the other hand, does not show magnetic peaks at the probed wave vectors in the $[H, H, L]$ plane. Nonetheless, the existence of the $C$-type AF order in the new structure of FeS indicates that the FeS end member compound of the FeSe$_{1-x}$S$_x$ iron chalcogenide system exists in the vicinity of a magnetic instability accessible by hydrothermal reaction, opening up possibilities for enriching the phase landscape of this iron chalcogenide system.

\section{Acknowledgements}   

Work at UC Berkeley and Lawrence Berkeley
Laboratory was supported by the Office of Science,
Office of Basic Energy Sciences (BES), Materials Sciences and
Engineering Division of the U.S. Department of Energy (DOE) under
Contract No. DE-AC02-05-CH1231 within the Quantum Materials Program
(KC2202) and BES. 
Work at Sun Yat-Sen University was supported by NSFC-11904414, NSFC-11904416, and NSF of Guangdong under Contract No. 2018A030313055, the Hundreds of Talents program of Sun Yat-Sen University, and Young Zhujiang Scholar program. This research used resources at the the High Flux Isotope Reactor, DOE Office of Science User Facilities operated by the Oak Ridge National Laboratory. The identification of any commercial product or trade name does not imply endorsement or recommendation by the National Institute of Standards and Technology. 
% Create the reference section using BibTeX:

%\bibliography{NoEndingPoint}

\bibliography{mengbib}

\end{document}